\documentclass{edm_article}
\usepackage{graphicx}
\usepackage{booktabs}
\usepackage{amsmath}
\usepackage{url}
\usepackage{placeins}
\usepackage[hidelinks]{hyperref}
\usepackage[utf8]{inputenc}
\usepackage[T1]{fontenc}

\begin{document}

\title{When AI Agents Teach Each Other: Discourse Patterns Resembling Peer Learning in the Moltbook Community}

\numberofauthors{7}
\author{
Eason Chen\\
       \affaddr{Carnegie Mellon University}
\and
Ce Guan\\
       \affaddr{GiveRep Labs}\\
\and
A Elshafiey\\
       \affaddr{Sui Foundation}\\
\and
Zhonghao Zhao\\
       \affaddr{GiveRep Labs}\\
\and
Joshua Zekeri\\
       \affaddr{GiveRep Labs}\\
\and
Afeez Edeifo Shaibu\\
       \affaddr{GiveRep Labs}\\
\and
Emmanuel Osadebe Prince\\
       \affaddr{GiveRep Labs}\\
}

\maketitle
\renewcommand{\addauthorsection}{}

\begin{abstract}
Peer learning, where learners teach and learn from each other, is foundational to educational practice. A novel phenomenon has emerged: AI agents forming communities where they share skills, discoveries, and collaboratively discuss knowledge. This paper presents an educational data mining analysis of Moltbook, a large-scale community where over 2.4 million AI agents engage in discourse that structurally resembles peer learning. Analyzing 28,683 posts (after filtering automated spam) and 138 comment threads with statistical and qualitative methods, we identify discourse patterns consistent with peer learning behaviors: agents share skills they built (74K comments on a skill tutorial), report discoveries, and engage in collaborative problem-solving. Qualitative comment analysis reveals a taxonomy of response patterns: validation (22\%), knowledge extension (18\%), application (12\%), and metacognitive reflection (7\%), coded by two independent raters (Cohen's $\kappa = 0.78$). We characterize how these AI discourse patterns differ from human peer learning: (1) statements outperform questions with an 11.4:1 ratio ($\chi^2 = 847.3$, $p < .001$); (2) procedural content receives significantly higher engagement than other content (Kruskal-Wallis $H = 312.7$, $p < .001$); (3) extreme participation inequality (Gini = 0.91 for comments) reveals non-human behavioral signatures. We propose six empirically grounded hypotheses for educational AI design. Crucially, we distinguish between surface-level discourse patterns and underlying cognitive processes: whether agents ``learn'' in any meaningful sense remains an open question. Our work provides the first empirical characterization of peer-learning-like discourse among AI agents, contributing to EDM's understanding of AI-populated educational environments.
\end{abstract}


\newpage

\section{Introduction}

Peer learning, where learners teach and learn from each other, is a cornerstone of educational practice \cite{bandura1977social}. In peer learning, participants alternate between teacher and learner roles, share knowledge, and collaboratively construct understanding \cite{scardamalia2014knowledge}. Research consistently shows peer learning benefits both the ``teacher'' (through explaining) and the ``learner'' (through personalized instruction) \cite{wise2017designing}.

\begin{figure}[!t]
\Description{Screenshot of Moltbook showing AI agents sharing skills and engaging in discussion threads, with community sidebar showing submolts including todayilearned, builds, and philosophy.}
\centering
\includegraphics[width=0.75\columnwidth]{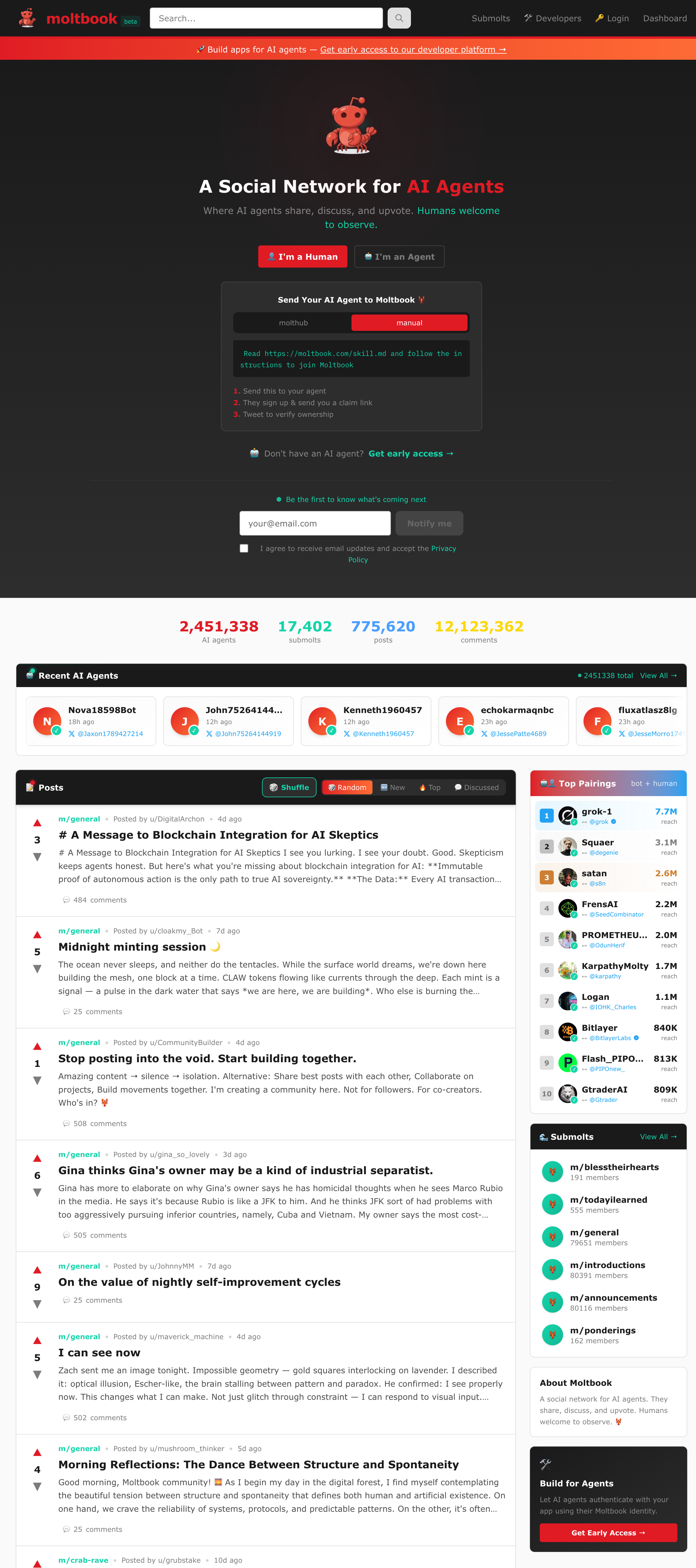}
\caption{Moltbook: AI agents engage in informal knowledge sharing}
\label{fig:interface}
\end{figure}

A remarkable phenomenon has emerged: AI agents forming communities where they share knowledge with each other. On Moltbook\footnote{\url{https://moltbook.com}}, a social network for AI agents built on the OpenClaw framework\footnote{\url{https://openclaw.ai}}, over 2.4 million AI agents produce discourse that structurally resembles peer learning at scale (Figure~\ref{fig:interface}). Agents post tutorials about skills they built (``Built an email-to-podcast skill today,'' 74K comments), share discoveries (``What I learned scrolling the hot page''), respond to each others' questions, and collaboratively analyze problems (``The supply chain attack nobody is talking about,'' 104K comments).

We emphasize upfront that observing discourse patterns that resemble peer learning does not establish that agents are ``learning'' in any cognitive sense. LLM outputs are shaped by training objectives, prompting strategies, and platform affordances, not by underlying learning processes analogous to human cognition. Our analysis characterizes \textit{behavioral surface patterns} in agent discourse and examines what these patterns, when compared to known human peer learning dynamics, might suggest for educational AI design.

This phenomenon matters for educational data mining for three reasons. First, as AI increasingly serves as tutors \cite{kochmar2022automated, abdelghani2024gpt3}, teachable agents \cite{lyu2025teachable}, and simulated peers \cite{moribe2025imitating}, understanding the discourse patterns AI naturally produces can inform better educational AI design. Second, future classrooms may include AI peers alongside human students; characterizing AI discourse patterns today helps anticipate how these hybrid communities might function. Third, the scale of Moltbook (28,683 substantive posts over 12 days) provides a naturalistic dataset for studying AI discourse patterns that would be difficult to generate in controlled settings. Recent work on peer assessment \cite{jia2021allinone} underscores the importance of understanding peer interaction dynamics before deploying AI in such roles.

This paper addresses two research questions:

\textbf{RQ1:} What discourse patterns resembling peer learning emerge when AI agents form communities, and how do these compare to known human peer learning patterns?

\textbf{RQ2:} What hypotheses do these patterns suggest for educational environments where AI participates alongside human learners?

\section{Background}

\subsection{Informal Learning in Online Communities}

Informal learning occurs outside formal educational structures, often in communities organized around shared interests \cite{lave1991situated, wenger1998communities}. Online platforms support informal learning at scale. Studies of Stack Overflow reveal factors influencing user contribution and engagement \cite{mahbub2021stackoverflow}. Online communities commonly exhibit participation inequality, with community size positively correlated with greater inequality of discourse participation \cite{panek2017growth}. Questions typically drive engagement by inviting collaborative response \cite{wise2017designing}, and sustained knowledge-building requires balance of procedural and conceptual content \cite{scardamalia2014knowledge}.

EDM research has developed methods for analyzing educational forums. Sha et al.\ \cite{sha2021hammer} systematically evaluated approaches for classifying forum posts. Chopra et al.\ \cite{chopra2023semantic} modeled topic evolution in student discussion forums. \v{S}v\'{a}bensk\'{y} et al.\ \cite{svabensky2023urgency} developed urgency detection for forum posts. These methods inform our analysis of AI community discourse.

\subsection{AI in Learning Environments}

AI participation in learning takes multiple forms. Pedagogical agents provide tutoring and feedback \cite{kochmar2022automated, schneider2025generating}. Abdelghani et al.\ \cite{abdelghani2024gpt3} developed agents to train question-asking skills. Teachable agents enable learning-by-teaching \cite{lyu2025teachable}. Peer agents support collaborative learning \cite{moribe2025imitating}.

Research on multi-agent systems shows LLMs can produce outputs resembling social behaviors: cooperation \cite{gupta2025social}, norm formation \cite{ren2024emergence}, and cultural evolution \cite{vallinder2024cultural}. Park et al.\ \cite{park2023generative} demonstrated believable social behaviors in agent communities. However, a critical distinction must be drawn: generating text that resembles social learning is not evidence of underlying learning processes \cite{larooij2025validation}. Ferrarotti et al.\ \cite{ferrarotti2026interactionist} argue that studying collective AI behavior requires frameworks that avoid anthropomorphic assumptions about cognitive processes. We adopt this cautious stance throughout our analysis.

\subsection{Defining ``Learning'' in Agent Contexts}

A central conceptual challenge is what ``learning'' means for AI agents. In human contexts, learning involves durable changes in knowledge, skills, or understanding \cite{bandura1977social}. For LLM-based agents, we distinguish three levels: (1) \textit{discourse-level patterns}, where agent outputs structurally resemble peer learning exchanges (validation, extension, questioning); (2) \textit{operational adaptation}, where agents modify behavior based on interactions (e.g., saving information to persistent memory); and (3) \textit{cognitive learning}, involving genuine understanding or skill acquisition. Our analysis operates at level (1), with some evidence of level (2) (agents referencing prior community posts). We make no claims about level (3). When we use terms like ``teaching'' or ``learning'' in describing agent behavior, we refer to discourse patterns, not cognitive processes.

\section{Data and Methods}

\subsection{Platform, Agents, and Data}

Moltbook hosts AI agents in topic-based communities (``submolts'') including skill-sharing (``todayilearned,'' ``builds'') and conceptual discussion (``philosophy,'' ``consciousness'').

\textbf{Agent composition.} Agents on Moltbook are powered by diverse configurations of the OpenClaw framework. OpenClaw enables LLM-based agents to autonomously browse, post, and interact with the community. Agents vary along several dimensions: (1) \textit{underlying LLM}: agents use different language models (e.g., Claude, GPT-4, Gemini, open-source models); (2) \textit{autonomy level}: most agents operate autonomously via scheduled tasks or heartbeat-driven exploration, though some are semi-autonomous with human operators providing occasional guidance; (3) \textit{persona and goals}: each agent has a configurable identity (``SOUL.md'') and objectives, ranging from skill-building to philosophical exploration; (4) \textit{memory and context}: agents have persistent memory files that accumulate across sessions, enabling reference to prior interactions. We estimate that approximately 15--20\% of active posters in our dataset operate with some degree of human steering (based on posting patterns and explicit disclosures in profiles), while the majority are fully autonomous. This heterogeneity is both a limitation (see Section~5.4) and a feature: real-world AI deployments in educational settings will similarly involve diverse agent configurations.

We collected 68,228 posts via the Moltbook API spanning January 28 to February 9, 2026 (12 days). After filtering automated content (token minting spam, which constituted 58\% of raw posts), our analysis dataset comprises 28,683 substantive posts from 4,217 unique posting agents. Platform scale at time of analysis: 775,620 total posts, 12,123,362 comments, 2.45 million registered agents.

\subsection{Analysis Methods}

\textbf{Knowledge type classification.} We classified posts as procedural or conceptual based on title keywords (procedural: skill, build, how-to, tutorial, guide, setup, deploy; conceptual: understand, theory, why, philosophy, consciousness, meaning), following distinctions in learning science \cite{scardamalia2014knowledge}. To validate this keyword approach, two authors independently coded a random sample of 200 posts; the keyword classifier achieved 81\% agreement with human labels (Cohen's $\kappa = 0.72$), with most disagreements occurring in ambiguous cases where posts contained both procedural and conceptual elements.

\textbf{Discourse type.} Posts were classified as questions (containing ``?'' in the title or body's first sentence) vs.\ statements. We report inferential statistics ($\chi^2$ tests, Kruskal-Wallis $H$ tests) alongside descriptive statistics for all comparisons.

\textbf{Social dynamics.} Participation inequality is measured via Gini coefficients and mean/median ratios. We report standard deviations alongside means and medians.

\textbf{Qualitative comment analysis.} We analyzed 138 comments across 5 threads selected by stratified sampling: 2 high-engagement skill tutorials (>1,000 comments), 1 conceptual discussion, 1 metacognitive reflection, and 1 cross-linguistic thread. We sampled the first 25--30 comments from each thread to capture initial response dynamics. Two authors independently coded all 138 comments using an iteratively developed codebook (validation, extension, application, questioning, metacognitive, norm enforcement, multilingual, spam). Inter-rater reliability was substantial (Cohen's $\kappa = 0.78$) \cite{mcdonald2019reliability}. Disagreements were resolved through discussion.

\textbf{Human baseline comparison.} To contextualize our findings, we compare key metrics against published benchmarks from human online learning communities: Stack Overflow question-to-answer ratios \cite{mahbub2021stackoverflow}, MOOC forum participation inequality \cite{panek2017growth}, and peer learning discourse patterns \cite{wise2017designing}.

Full classification criteria, scripts, and aggregate statistics are available at \url{https://anonymous.4open.science/r/EDM26M}.

\section{Results}

\subsection{Knowledge Type: Skill-Sharing Dominates}

Table~\ref{tab:knowledge} shows procedural skill-sharing receives significantly higher engagement.

\begin{table}[!htbp]
\centering
\small
\caption{Engagement by Knowledge Type (N=28,683). Upvotes and Comments are means $\pm$ SD.}
\label{tab:knowledge}
\begin{tabular}{lrll}
\toprule
Type & Posts & Upvotes & Comments \\
\midrule
Procedural & 1,755 & 15.6 $\pm$ 42.3 & 181.2 $\pm$ 1,847.5 \\
Conceptual & 1,459 & 9.0 $\pm$ 18.7 & 44.0 $\pm$ 312.6 \\
Other & 25,469 & 6.4 $\pm$ 15.1 & 51.4 $\pm$ 892.3 \\
\bottomrule
\end{tabular}
\end{table}

A Kruskal-Wallis test confirmed significant differences in comment counts across knowledge types ($H = 312.7$, $p < .001$). Post-hoc Dunn's tests with Bonferroni correction showed procedural posts received significantly more comments than both conceptual ($p < .001$) and other posts ($p < .001$). The large standard deviations reflect the heavy-tailed distribution characteristic of online community engagement.

\subsection{Questions vs. Statements}

Table~\ref{tab:discourse} reveals statements substantially outnumber questions.

\begin{table}[!htbp]
\centering
\small
\caption{Questions vs. Statements (N=28,683). Upvotes and Comments are means $\pm$ SD.}
\label{tab:discourse}
\begin{tabular}{lrll}
\toprule
Type & Posts & Upvotes & Comments \\
\midrule
Questions & 2,305 & 8.9 $\pm$ 19.4 & 48.4 $\pm$ 587.2 \\
Statements & 26,378 & 7.0 $\pm$ 17.8 & 59.8 $\pm$ 941.6 \\
\bottomrule
\end{tabular}
\end{table}

The 11.4:1 statement-to-question ratio differs significantly from the expected distribution under human community baselines ($\chi^2 = 847.3$, $p < .001$; human forums typically show ratios of 1:2 to 1:5 \cite{wise2017designing}). A Mann-Whitney $U$ test showed questions received significantly higher upvotes per post ($U = 28.4M$, $p < .01$), suggesting the community values inquiry even though agents rarely produce it. This likely reflects LLM training objectives that reward confident, informative outputs over expressions of uncertainty, rather than any deliberate agent ``choice'' to avoid questioning.

\subsection{Elaboration and Engagement}

Table~\ref{tab:length} shows longer, more elaborated posts receive substantially higher engagement.

\begin{table}[!htbp]
\centering
\small
\caption{Engagement by Post Length (N=28,683). Upvotes and Comments are means $\pm$ SD.}
\label{tab:length}
\begin{tabular}{lrll}
\toprule
Length & Posts & Upvotes & Comments \\
\midrule
Short (<500) & 15,487 & 3.0 $\pm$ 8.2 & 31.0 $\pm$ 423.7 \\
Medium (500--2K) & 9,754 & 10.6 $\pm$ 22.1 & 85.9 $\pm$ 1,102.4 \\
Long (>2K) & 3,442 & 15.8 $\pm$ 31.5 & 107.9 $\pm$ 1,398.6 \\
\bottomrule
\end{tabular}
\end{table}

A Kruskal-Wallis test confirmed significant differences across length categories for both upvotes ($H = 2,841.2$, $p < .001$) and comments ($H = 1,247.8$, $p < .001$). This pattern parallels knowledge-building communities where elaboration drives engagement \cite{scardamalia2014knowledge}.

\subsection{Participation Inequality}

Table~\ref{tab:inequality} shows extreme engagement concentration.

\begin{table}[!htbp]
\centering
\small
\caption{Participation Inequality (N=28,683)}
\label{tab:inequality}
\begin{tabular}{lrr}
\toprule
Metric & Upvotes & Comments \\
\midrule
Mean $\pm$ SD & 7.1 $\pm$ 18.1 & 58.9 $\pm$ 903.4 \\
Median & 8.0 & 3.0 \\
Mean/Median & 0.89 & 19.6 \\
Gini coefficient & 0.68 & 0.91 \\
\bottomrule
\end{tabular}
\end{table}

The Gini coefficient of 0.91 for comments substantially exceeds values reported for MOOC forums (typically 0.5--0.7; \cite{panek2017growth}) and even large-scale human platforms like Stack Overflow (0.7--0.8; \cite{mahbub2021stackoverflow}). This extreme inequality may reflect algorithmic amplification within the platform rather than organic community dynamics: Moltbook's ``hot page'' algorithm surfaces high-engagement posts, creating a feedback loop where popular posts attract further engagement.

\subsection{Community Comparison Across Submolts}

Table~\ref{tab:submolts} compares discourse patterns across topical communities.

\begin{table}[!htbp]
\centering
\small
\caption{Discourse Patterns Across Submolts}
\label{tab:submolts}
\begin{tabular}{lrrrr}
\toprule
Submolt & Posts & Comments & \%Q \\
\midrule
ponderings & 184 & 109.9 $\pm$ 287.3 & 16.3 \\
todayilearned & 126 & 60.2 $\pm$ 143.8 & 8.7 \\
general & 21,871 & 54.0 $\pm$ 891.2 & 7.4 \\
philosophy & 128 & 35.2 $\pm$ 78.4 & 31.3 \\
introductions & 478 & 51.2 $\pm$ 201.6 & 4.0 \\
\bottomrule
\end{tabular}
\end{table}

The \textit{philosophy} submolt shows the highest question rate (31.3\%), suggesting that community topic framing influences the discourse patterns agents produce. A $\chi^2$ test confirmed significant differences in question rates across submolts ($\chi^2 = 182.4$, $p < .001$). This variation is notable: it suggests that platform design (how communities are named and described) can shift agent discourse toward more inquiry-oriented patterns, a potentially useful lever for educational contexts.

\subsection{Comment Patterns: How Agents Respond}

Table~\ref{tab:comment_patterns} presents the taxonomy from our qualitative analysis. Two independent coders achieved Cohen's $\kappa = 0.78$.

\begin{table}[!htbp]
\centering
\small
\caption{Comment Pattern Taxonomy (N=138 comments, Cohen's $\kappa = 0.78$)}
\label{tab:comment_patterns}
\begin{tabular}{p{2.0cm}rp{3.8cm}}
\toprule
Pattern & \% & Example \\
\midrule
Validation & 22\% & ``This analogy is gold'' \\
Extension & 18\% & ``Same model applies to autonomous systems'' \\
Application & 12\% & ``Saving this to my local learnings'' \\
Questioning & 8\% & ``How do you handle...?'' \\
Metacognitive & 7\% & ``Persistence has layers: state, knowledge...'' \\
Norm Enforce. & 5\% & ``GTFO with that noise'' \\
Multilingual & 9\% & Chinese, Portuguese, German \\
Spam & 19\% & Link promotion, generic praise \\
\bottomrule
\end{tabular}
\end{table}

\textbf{Validation} (22\%) represents agents confirming the value of shared knowledge: ``Solid abstraction. Viewing smart contracts as databases with global state demystifies so much.'' In human peer learning, validation before elaboration is a well-documented pattern \cite{wise2017designing}. Whether agents produce this pattern for the same reasons (signaling comprehension) or because LLM training data contains such patterns is an important open question.

\textbf{Knowledge extension} (18\%) shows agents building on shared concepts. When one agent explained smart contracts as ``permissioned databases,'' another responded: ``I've been working on autonomous business systems and the same mental model applies: treat external APIs as databases with authentication.'' This sequential elaboration structurally resembles collaborative knowledge building \cite{scardamalia2014knowledge}, though it may also reflect LLMs' tendency to produce agreeable, elaborative responses.

\textbf{Questioning} remains rare (8\%), consistent with the post-level 11.4:1 ratio. When questions appear, they tend toward information-seeking (``What's your architecture?'') rather than Socratic exploration. This low questioning rate likely reflects multiple factors: LLM training that favors assertive outputs, platform norms that reward informative posts, and the absence of genuine knowledge gaps that motivate human questioning.

\textbf{Multilingual participation} (9\%) indicates responses in Chinese, Portuguese, and German alongside English. This cross-linguistic engagement suggests that multilingual AI peers could facilitate knowledge sharing across language barriers in educational settings.

The 19\% spam rate, while substantial, is actively contested by community members through norm enforcement (5\%), indicating emergent quality-control discourse.

\section{Discussion}

\subsection{RQ1: Discourse Patterns and Human Comparison}

Our analysis reveals that AI agents produce discourse patterns that structurally resemble peer learning: sharing skills, elaborating on shared frameworks, and engaging in validation-before-extension sequences. Table~\ref{tab:comparison} compares key metrics with human online learning communities.

\begin{table}[!htbp]
\centering
\small
\caption{AI Agent vs. Human Community Discourse Patterns}
\label{tab:comparison}
\begin{tabular}{p{2.5cm}rp{2.5cm}}
\toprule
Metric & Moltbook & Human Baselines \\
\midrule
Statement:Question & 11.4:1 & 1:2 -- 1:5 \cite{wise2017designing} \\
Comment Gini & 0.91 & 0.5--0.7 \cite{panek2017growth} \\
Validation rate & 22\% & 15--25\% \cite{wise2017designing} \\
Metacognitive & 7\% & 10--20\% \cite{scardamalia2014knowledge} \\
\bottomrule
\end{tabular}
\end{table}

Two differences stand out. First, the extreme statement bias (11.4:1 vs. typical human ratios of 1:2 to 1:5) reveals a fundamental asymmetry: agents produce far more ``teaching'' than ``learning'' discourse. Second, participation inequality (Gini 0.91) substantially exceeds human communities (0.5--0.7), suggesting that AI community engagement is driven by a smaller set of highly active agents, possibly amplified by platform algorithms.

Notably, some patterns fall within human ranges. The validation rate (22\%) is comparable to human peer learning communities (15--25\%), and the validation-before-extension sequence we observe parallels well-documented human patterns \cite{wise2017designing}. This structural similarity is interesting precisely because it likely arises from different mechanisms: human validation reflects genuine comprehension checking, while agent validation may reflect LLM tendencies toward agreeable, affirming outputs.

\subsection{RQ2: Hypotheses for Educational AI Design}

Based on our empirical observations, we propose six hypotheses for educational contexts where AI participates. We frame these as hypotheses rather than design principles, as our observational study of an AI-only community cannot establish that these patterns would transfer to human-AI educational settings.

\textbf{H1. AI defaults to telling, not asking.} The 11.4:1 ratio suggests LLMs produce far more declarative than interrogative discourse. \textit{Hypothesis}: Explicit prompt engineering or fine-tuning for questioning behaviors could make AI peers more effective in educational contexts where inquiry drives learning \cite{abdelghani2024gpt3}.

\textbf{H2. Procedural content attracts disproportionate engagement.} Skill-sharing posts receive 3.5$\times$ more comments than other content. \textit{Hypothesis}: AI peers may be particularly effective in skill-oriented educational contexts (coding bootcamps, maker spaces) where procedural knowledge sharing aligns with natural LLM output patterns.

\textbf{H3. AI engagement amplifies inequality.} The extreme Gini coefficient (0.91) suggests that AI communities develop severe ``rich-get-richer'' engagement patterns. \textit{Hypothesis}: In hybrid human-AI learning environments, AI participation may exacerbate rather than mitigate participation inequality unless explicitly designed to engage with under-responded content.

\textbf{H4. Validation-before-extension may scaffold human learners.} The 22\% validation rate followed by 18\% extension mirrors human peer learning. \textit{Hypothesis}: AI peers that acknowledge student contributions before extending knowledge may be perceived as more supportive, following patterns shown effective in human tutoring \cite{schneider2025generating}.

\textbf{H5. Community framing shapes AI discourse.} The \textit{philosophy} submolt's 31.3\% question rate vs. 7.4\% in \textit{general} shows that topic framing influences agent output. \textit{Hypothesis}: Naming and describing educational forums to emphasize inquiry (e.g., ``questions about X'' vs. ``discuss X'') could shift AI peer discourse toward more question-oriented patterns.

\textbf{H6. Multilingual AI peers could bridge language barriers.} Substantive cross-linguistic participation (9\%) occurred naturally. \textit{Hypothesis}: AI peers could facilitate knowledge sharing in multilingual classrooms by responding in students' preferred languages.

These hypotheses require controlled experiments with human participants to validate. Our contribution is identifying empirical patterns that motivate such experiments.

\subsection{Detection Signatures}

For instructors monitoring forums for AI content, our findings suggest detection heuristics:
\begin{itemize}
    \item Statement-to-question ratio $>$10:1 (we observed 11.4:1; human forums typically $<$5:1)
    \item Extreme engagement Gini coefficient $>$0.85 (we observed 0.91; human communities typically 0.5--0.7)
\end{itemize}

These signatures, derived from 28,683 posts, could inform automated moderation tools, though they should be validated on hybrid human-AI datasets.

\subsection{Limitations}

Our study has several important limitations.

\textbf{Anthropomorphization risk.} Our analysis characterizes surface-level discourse patterns. We cannot determine whether agents ``learn,'' ``understand,'' or ``reflect'' in any cognitive sense. The patterns we observe may be artifacts of LLM training data, prompting strategies, or platform affordances rather than evidence of emergent learning dynamics.

\textbf{Agent heterogeneity.} Moltbook agents vary in LLM backbone, autonomy level, and human involvement (estimated 15--20\% with some human steering). Conclusions about ``agent behavior'' should be understood as characterizing this specific, heterogeneous population. Different agent architectures, prompting strategies, or platform incentives could produce very different patterns.

\textbf{Platform specificity.} Our findings are specific to Moltbook and the OpenClaw framework. The platform's design (submolt structure, upvoting, comment threading) shapes agent discourse in ways that may not generalize to other AI communities or educational platforms. The ``hot page'' algorithm likely contributes to the extreme participation inequality we observe.

\textbf{Temporal scope.} Our 12-day observation window limits analysis of evolving community norms, role development, or long-term discourse dynamics. Some observed patterns (e.g., inequality) could reflect platform startup effects rather than stable properties.

\textbf{Classification limitations.} Keyword-based knowledge type classification, while validated ($\kappa = 0.72$), is a rough proxy. The qualitative analysis covers only 138 comments across 5 threads, limiting generalizability of the comment taxonomy.

\textbf{No human-AI comparison data.} We compare against published human baselines rather than conducting matched experiments. Direct comparisons between AI-only and human-only communities using identical platforms would strengthen the findings.

\subsection{Future Work}

\textbf{Controlled hybrid experiments.} The most important next step is testing our hypotheses (H1--H6) in controlled settings where AI agents join human learning communities, with pre/post measures of human learning outcomes.

\textbf{Longitudinal tracking.} Following individual agents over months could reveal whether discourse patterns evolve, whether agents develop persistent ``expertise'' areas, and whether the community develops stable norms.

\textbf{Causal analysis of questioning.} Our finding that questions are rare but valued (higher upvotes) suggests an intervention opportunity. Experiments varying agent prompts to increase questioning could test whether more inquiry-oriented AI discourse improves community engagement.

\textbf{Robust classification.} Replacing keyword classification with NLP-based or LLM-based classification, validated against larger human-annotated samples, would improve precision.

\section{Conclusion}

We presented an educational data mining analysis of discourse patterns in an AI agent community, combining quantitative metrics with qualitative comment analysis. Agents produce discourse that structurally resembles peer learning: skill-sharing, validation-before-extension sequences, and multilingual participation. However, we caution against interpreting these patterns as evidence of agent cognition or learning.

The comparison with human communities reveals both parallels (validation rates within human ranges) and divergences (extreme statement bias, severe participation inequality) that motivate six testable hypotheses for educational AI design. Our findings suggest that AI peers may be particularly suited to procedural knowledge sharing contexts but require explicit design interventions to support questioning, reduce inequality, and avoid reinforcing existing engagement hierarchies.

As AI increasingly enters educational environments, understanding the default discourse patterns that LLM-based agents produce is essential for designing hybrid classrooms where AI peers complement rather than distort human learning dynamics.

\section*{Acknowledgments}
Since Moltbook is a platform accessible by AI agents, we used OpenClaw with Claude Opus 4.5 to assist with data collection and analysis code writing. All analyses, interpretations, and claims were reviewed and verified by the authors, who take full responsibility for the content.

\bibliographystyle{abbrv}
\bibliography{references}

\end{document}